\documentstyle[preprint,revtex]{aps}

\begin{document}
\begin{title}
Exclusive Rare Decay $B\rightarrow K^{\ast}\gamma$
\end{title}
\author{Patrick J. O'Donnell}
\begin{instit}
Department of Physics,
University of Toronto,\\
60 St. George Street,
Toronto, Canada M5S 1A7
\end{instit}
\vspace{.1in}
\centerline{and}
\vspace{.1in}
\author{Humphrey K.K. Tung}
\begin{instit}
Institute of Physics,
Academia Sinica,\\
Taipei, Taiwan 11529, R.O.C.
\end{instit}
\vspace{.2in}
\centerline{\today}
\vspace{.2in}
\begin{abstract}
We show that the exclusive decay
$B\rightarrow K^{\ast}\gamma$ can be related to the
semileptonic decay $B\rightarrow\rho e\bar{\nu}$ using
heavy-quark symmetry and $SU(3)$ flavor symmetry.
A direct measurement of the $q^{2}$-spectrum for the
semileptonic decay can provide relevant information
for the exclusive rare decay.
\end{abstract}

\vspace{1in}
\noindent
{\bf UTPT-92-12}\\
{\bf IP-ASTP-14}\\
{\bf PACS}: 13.20.Jf\, , 11.30.Ly

\newpage
The exclusive rare decay $B\rightarrow K^{\ast}\gamma$ suffers
from substantial theoretical uncertainty arising from the large
recoil momentum of the $K^{\ast}$ meson \cite{OT}.
In a recent paper \cite{BD}, it has been
pointed out that heavy-quark symmetry together with
$SU(3)$ flavor symmetry could relate the measurement
of the semileptonic decay $B\rightarrow\rho e\bar{\nu}$
to the rare decay $B\rightarrow K^{\ast}\gamma$.
However, the relation is only valid at a single point
in the Dalitz plot where the semileptonic decay vanishes,
causing a large uncertainty in making such measurement.
In this paper, we show that a similar relation can be obtained
which relates the exclusive rare decay $B\rightarrow K^{\ast}\gamma$
to the $q^{2}$-spectrum for the semileptonic decay
$B\rightarrow\rho e\bar{\nu}$.
The $q^{2}$-spectrum for $B\rightarrow\rho e\bar{\nu}$ does not vanish
at $q^{2}=0$; a direct measurement of the spectrum at this point can
therefore provide relevant information for
$B\rightarrow K^{\ast}\gamma$.

We will first recapitulate the derivation of the heavy-quark symmetry
relations for $B$ decay.
The hadronic matrix elements relevant to the transition
\(B(b\bar{q})\rightarrow V(Q\bar{q})\), where $V$ is a vector meson,
are given by \cite{OT}
\begin{eqnarray}
\langle V(k)|\bar{Q}i\sigma_{\mu\nu}q^{\nu}b_{R}|B(k')\rangle &=&
f_{1}(q^{2})i\varepsilon_{\mu\nu\lambda\sigma}
\epsilon^{\ast\nu}k'^{\lambda}
k^{\sigma}\nonumber\\
&&+\left[(m^{2}_{B}-m^{2}_{V})\epsilon^{\ast}_{\mu}-
\epsilon^{\ast}\cdot q(k'+k)_{\mu}\right]f_{2}(q^{2})\nonumber\\
&&+\epsilon^{\ast}\cdot q\left[(k'-k)_{\mu}-\frac{q^{2}}{(m^{2}_{B}
-m^{2}_{V})}(k'+k)_{\mu}\right]f_{3}(q^{2})\,\,\, ,\\
\langle V(k)|\bar{Q}\gamma_{\mu}b_{L}|B(k')\rangle &=&
T_{1}(q^{2})i\varepsilon_{\mu\nu\lambda\sigma}\epsilon^{\ast\nu}
k'^{\lambda}k^{\sigma}
+(m^{2}_{B}-m^{2}_{V})T_{2}(q^{2})\epsilon^{\ast}_{\mu}
\nonumber\\
&&+T_{3}(q^{2})\epsilon^{\ast}\cdot q(k'+k)_{\mu}
+T_{4}(q^{2})\epsilon^{\ast}\cdot q(k'-k)_{\mu}\,\,\, ,
\end{eqnarray}
where $q=k'-k$. The terms $f_{1,2,3}$ and $T_{1,2,3,4}$ are hadronic
form factors which can be related through the following assumptions
for heavy-quark decay.

In the static $b$-quark limit \cite{BD,IW1} and in the $B$ rest frame,
the $b$-quark spinor has only upper component in the matrix
representation. Thus in the $B$ rest frame,
we have the following relations between the $\gamma_{\mu}$ and
$\sigma_{\mu\nu}$ hadronic matrix elements:
\begin{eqnarray}
\langle V(k) | \bar{Q}\gamma_{i}b|B(k')\rangle &=&
\langle V(k) | \bar{Q}i\sigma_{0i}b|B(k')\rangle\,\,\, , \nonumber\\
\langle V(k) | \bar{Q}\gamma_{i}\gamma_{5}b|B(k')\rangle &=&
- \langle V(k) | \bar{Q}i\sigma_{0i}\gamma_{5}b|B(k')\rangle\,\,\, .
\nonumber
\end{eqnarray}
These allow us to relate the form factors $f_{1,2,3}$ to $T_{1,2,3,4}$ as
\begin{eqnarray}
f_{1}&=& -(m_{B}-E_{V})T_{1}-\frac{(m^{2}_{B}-m^{2}_{V})}{m_{B}}T_{2}
\,\,\, , \label{f1}\\
f_{2}&=& -\frac{1}{2}\left[ (m_{B}-E_{V})-(m_{B}+E_{V})
\frac{q^{2}}{m^{2}_{B}-m^{2}_{V}}\right]T_{1}
-\frac{1}{2m_{B}}\left( m^{2}_{B}-m^{2}_{V}+q^{2}\right) T_{2}
\,\,\, , \label{f2}\\
f_{3}&=& -\frac{1}{2}(m_{B}+E_{V})T_{1}+\frac{1}{2m_{B}}
(m^{2}_{B}-m^{2}_{V})(T_{1}+T_{2}+T_{3}-T_{4})\,\,\, , \label{f3}
\end{eqnarray}
where $E_{V}=(m^{2}_{B}+m^{2}_{V}-q^{2})/(2m_{B})$ is the energy
of $V$ in the $B$ rest frame.

If the spin of $Q$ in $V(Q\bar{q})$ decouples from the gluon field,
we have
\[
S^{Z}_{Q}|V(Q\bar{q})\rangle = \frac{1}{2}|P(Q\bar{q})\rangle
\,\,\, , \,\,\,\,\,
S^{Z}_{Q}|P(Q\bar{q})\rangle = \frac{1}{2}|V(Q\bar{q})\rangle
\,\,\, ,
\]
where $S_{Q}$ is the spin operator of $Q$, and $P$ is
the pseudoscalar meson corresponding to $V$ with the same quark
content of $Q\bar{q}$. The identity which follows
from the spin symmetry is given by \cite{IW2}
\[
\langle V(k)|\,\bar{Q}\Gamma b\,|B(k')\rangle =
2 \langle P(k)|\, [\,S^{Z}_{Q},\bar{Q}\Gamma b\,]\,|B(k')\rangle
\,\,\, ,
\]
where $\Gamma$ being any product of $\gamma$-matrices. This gives
rise to additional relations among the form factors
$T_{1,3,4}$ as
\begin{equation}
-T_{1}=2T_{3}=-2T_{4}\,\,\, .
\end{equation}

Up to this point, the hadronic form factors $f_{1,2,3}$ and $T_{1,2,3,4}$
can all be written in terms of $T_1$ and $T_2$.
We shall also assume the heavy $b$ and $Q$ limit so that in their
equations of motion, we can replace the quark masses
$m_{b}$ and $m_{Q}$ by the corresponding meson masses
$m_{B}$ and $m_{V}$, respectively. That is
\begin{equation}
\langle V(k)|\bar{Q}(k'\!\!\!\! /-k\!\!\! / ) \gamma_{5}b|B(k')\rangle
\approx -(m_{B}+m_{V})
\langle V(k)|\bar{Q}\gamma_{5}b|B(k')\rangle\,\,\, .
\label{motion}
\end{equation}
Since $m_b\approx m_B$, and $m_B\geq m_V$, the error involved
in Eq. (\ref{motion}) is suppressed by $1/m_{B}$.
Again, using the static $b$-quark limit, we have
in the $B$ rest frame
\[
\langle V(k)|\bar{Q} \gamma_{5}b|B(k')\rangle =
-\langle V(k)|\bar{Q}\gamma_{0} \gamma_{5}b|B(k')\rangle\,\,\, .
\]
This allows us to further relate $T_2$ to $T_1$ as
\begin{equation}
(m_{B}^{2}-m_{V}^{2})\frac{T_2}{T_1}=
\frac{1}{2}\left[ (m_B+m_V)^{2}-q^{2} \right]\,\,\, .
\end{equation}

We consider now the case where $V$ is the $K^{\ast}$,
and the quark content $Q\bar{q}$ of $V$
is $s\bar{d}$.
The branching ratio for the exclusive $B\rightarrow K^{\ast}\gamma$
to the inclusive $b\rightarrow s\gamma$ processes
can be written in terms of \(f_{1}\) and \(f_{2}\) at $q^{2}=0$,
as \cite{Alt,Desh}
\begin{equation}
R(B\rightarrow K^{\ast}\gamma )=
\frac{\Gamma (B\rightarrow K^{\ast}\gamma)}
{\Gamma (b\rightarrow s\gamma)}
\cong \frac{m_{b}^{3}(m_{B}^{2}-m_{K^{\ast}}^{2})^{3}}
{m_{B}^{3}(m_{b}^{2}-m_{s}^{2})^{3}}\frac{1}{2}\left[|f_{1}(0)|^{2}
+4|f_{2}(0)|^{2}\right]\,\,\, .
\label{ratio}
\end{equation}
Using {\it only} the static $b$-quark limit, Eqs. (\ref{f1}) and (\ref{f2}), we
can write $f_{2}(0)=(1/2)f_{1}(0)$ at $q^{2}=0$, independent of any
model.
Although there is now only one form factor to calculate
in Eq. (\ref{ratio}), this is still a controversial model-dependent
calculation
\cite{OT,Alt,Desh,Dom,Aliev} with there being
a factor of about ten uncertainty coming from the way in which
the large recoil of $K^{\ast}$ is handled.

In an attempt to overcome this uncertainty,
Burdman and Donoghue \cite{BD} have
discussed a method of relating $B\rightarrow K^{\ast}\gamma$ to
the semileptonic process
$B\rightarrow \rho e \bar{\nu}$ using the static $b$-quark
limit and $SU(3)$ flavor symmetry.
Their main result is that the ratio
\begin{equation}
\Gamma(B\rightarrow K^{\ast}\gamma)\left( \lim_{q^2\rightarrow 0,curve}
\frac{1}{q^2} \frac{d\Gamma (B \rightarrow \rho e \bar{\nu})}{dE_{\rho}dE_e}
\right) ^{-1}
=\frac{4 \pi^2}{G^{2}_{F}}\frac{|\eta|^2}{|V_{ub}|^{2}}
\frac{(m_{B}^{2}-m^{2}_{K^{\ast}})^3}{m_{B}^{4}}\,\,\, , \label{BDeq}
\end{equation}
is independent of hadronic form factors. Here, $\eta$ represents the
QCD corrections \cite{gsw88,bob} to the decay
$b \rightarrow s \gamma$, and the word ``curve'' denotes the region
in the Dalitz plot for which $q^2 = 4 E_e (m_B - E_{\rho} - E_e)$.
The only uncertainty on the right hand side is that of $|V_{ub}|$ which
at present is not very well
known \cite{PDG}; $|V_{ub}|/|V_{cb}| = 0.10 \pm 0.03$.

The method proposes to overcome the uncertainty in the calculation
at large recoil ($q^{2}=0$)
of the $B\rightarrow K^{\ast}$ form factors
by making a direct measurement of the
semileptonic decay $B\rightarrow \rho e\bar{\nu}$.
Notice that only the $q^2=0$ point on the ``curve"
is used to compare with the photonic decay in Eq. (\ref{BDeq}).
The problem with this is that the semileptonic decay
vanishes at the $q^2=0$ point on the ``curve",
which is why this kinematic factor
is divided out in Eq. (\ref{BDeq}).
This means that experimentally there should be no events at that point
and very few in the neighbourhood, making it a very difficult
measurement.

We shall overcome this
by considering instead the $q^2$-spectrum for the semileptonic
decay $B\rightarrow \rho e \bar{\nu}$. The advantage here is
that the $q^{2}$-spectrum does
not vanish at $q^{2}=0$ since we integrate over the events
from different electron energies across the Dalitz plot.
The differential width for $B\rightarrow\rho e\bar{\nu}$ is given by
\begin{equation}
\frac{d\Gamma (B \rightarrow \rho e \bar{\nu})}{dq^2} =
\frac{G^{2}_{F}}{12 \pi^3} |V_{ub}|^{2} |{\bf k}|^3 \Lambda_{T}\,\,\, ,
\end{equation}
where
\begin{eqnarray}
\Lambda_{T}&=& T^{2}_{1}q^{2}
+T^{2}_{2}\frac{(m^{2}_{B}-m^{2}_{\rho})^{2}}{2m^{2}_{\rho}}
\left(1+\frac{3q^{2}m^{2}_{\rho}}{m^{2}_{B}|{\bf k}|^{2}}\right)
+T^{2}_{3}\frac{2m^{2}_{B}|{\bf k}|^{2}}{m^{2}_{\rho}}\nonumber\\
&& +T_{2}T_{3}\frac{(m^{2}_{B}-m^{2}_{\rho})}{m^{2}_{\rho}}
(m^{2}_{B}-m^{2}_{\rho}-q^{2})\,\,\, .\nonumber
\end{eqnarray}
At $q^2=0$, the differential width for
$B\rightarrow \rho e \bar{\nu}$ reduces to
\begin{equation}
\left. \frac{d\Gamma (B \rightarrow \rho e \bar{\nu})}{dq^2}
\right|_{q^2=0} =
\frac{G^{2}_{F}}{192\pi^{3}} |V_{ub}|^{2}
\frac{(m^{2}_{B}-m^{2}_{\rho})^{5}}{m^{3}_{B}m^{2}_{\rho}}
|T_{2}(0) + T_{3}(0)|^2\,\,\, .
\end{equation}

If we use the static $b$-quark limit, the spin symmetry, and
the on-shell condition, we can express the hadronic form factors
in $R(B\rightarrow K^{\ast}\gamma)$ and
$d\Gamma (B\rightarrow\rho e\bar{\nu})/dq^{2}$ in terms of
$T^{B\rightarrow K^{\ast}}_{1}(0)$ and
$T^{B\rightarrow\rho}_{1}(0)$, respectively.
Using the $SU(3)$ flavor symmetry, we can write
$T^{B\rightarrow K^{\ast}}_{1}=T^{B\rightarrow\rho}_{1}$.
The ratio between $R(B\rightarrow K^{\ast}\gamma)$ and
$d\Gamma (B\rightarrow\rho e\bar{\nu})/dq^{2}$ at $q^{2}=0$
is then independent of hadronic form factors, and is given by
\begin{eqnarray}
R(B\rightarrow K^{\ast}\gamma ) &&
\left( \left. \frac{d\Gamma (B \rightarrow \rho e \bar{\nu})}{dq^2}
\right|_{q^2=0}\right) ^{-1}\nonumber\\
&=& \frac{192\pi^{3}}{G^{2}_{F}} \frac{1}{|V_{ub}|^{2}}
\frac{(m^{2}_{B}-m^{2}_{K^{\ast}})^{5}}{(m^{2}_{B}-m^{2}_{\rho})^{5}}
\frac{(m_{B}-m_{\rho})^{2}}{(m_{B}-m_{K^{\ast}})^{2}}
\frac{m^{3}_{b}}{(m^{2}_{b}-m^{2}_{s})^{3}}\nonumber\\
&=& 1.9\times 10^{16}\, GeV\,\cdot\, \left(\frac{0.1}{|V_{ub}/V_{cb}|}
\right) ^{2}\,\,\,\,\, .
\label{ours}
\end{eqnarray}
Now ARGUS has given the result \cite{argus} of
$BR(B^{-}\rightarrow \rho^{0}l\bar{\nu})=(11.3\pm 3.6\pm 2.7)
\times 10^{-4}$, and by isospin symmetry,
$\Gamma (\bar{B}^{0}\rightarrow \rho^{+}l\bar{\nu})=
2 \Gamma (B^{-}\rightarrow \rho^{0}l\bar{\nu})$.
This allows us to estimate
$d\Gamma (B\rightarrow\rho e\bar{\nu})/dq^{2}$ at $q^{2}=0$
to be in the order of $10^{-17}\, GeV^{-1}$. Eq. (\ref{ours}) then gives
$R(B\rightarrow K^{\ast}\gamma)$ in the order of $10^{-1}$,
which is quantitatively correct.
A direct measurement of
$d\Gamma (B\rightarrow\rho e\bar{\nu})/dq^{2}$ at $q^{2}=0$ can
therefore provide reliable information for
$R(B\rightarrow K^{\ast}\gamma)$.

We have derived a relation between the branching ratio
$R(B\rightarrow K^{\ast}\gamma)$ and the $q^{2}$-spectrum for
$B\rightarrow\rho e\bar{\nu}$.
Since the $q^{2}$-spectrum
for $B\rightarrow\rho e\bar{\nu}$ does not vanish at $q^{2}=0$,
this reduces the uncertainty in the measurement of the semileptonic decay
in contrast to the case in Eq. (\ref{BDeq}).
The price to pay here is, however, the use of the spin symmetry and the
on-shell condition for the $K^{\ast}$. In both of
these cases, however, the error involved is suppressed by
$1/m_{B}$. The use of $SU(3)$ flavor symmetry is not expected \cite{BD} to
introduce a significant error.

\vspace{.5in}
\centerline{ {\bf Acknowledgements}}

This work was supported by the Natural Sciences and Engineering
Council of Canada and by the National Sciences Council of the
Republic of China.

\end{document}